\documentstyle[editedbook,epsfig,psfig,epsf]{mq}

\def\etal{{\em et al.} }

\def\cm2{cm$^2$ }
\def\se1{s$^{-1}$ }

\def\grs{GRS 1915+105} 

\begin{opening}
\title{Comptonization, the X-ray-radio correlation and the long-term periodicity in the $\chi$-states of \grs}
\author{A. Rau$^{1,2}$ \& J. Greiner$^{1,2}$}
\institute{$^1$ Astrophysical Institute Potsdam, An der Sternwarte 16, 14482 Potsdam, Germany.\\
$^2$ Max-Planck-Institute for extraterrestrial Physics, Giessenbachstrasse, 85748 Garching, Germany.}
\end{opening}

\runningtitle{Analysis of the $\chi$-states of \grs}
\runningauthor{Rau \& Greiner}

\begin{document}
\vspace{-0.5cm}
\begin{abstract}
{\small We analyzed 139 $\chi$-state observations of \grs\ with {\it RXTE} from 1997 to 2000 and found i) that the observations fall into two groups with different Comptonization behavior, ii) that the slope of the hard X-ray component correlates with the radio flux, thus revealing the interaction of jet and corona, and iii) a 590\,days long term periodicity in the hard X-ray and radio components.}
\end{abstract}

\vspace{-0.4cm}
\section{Introduction}
\vspace{-0.1cm}
The prototypical microquasar \grs\ was discovered by {\it Granat} \cite{cbl92} as a transient X-ray source. {\it RXTE} observations revealed astonishing X-ray variability \cite{gmr96} which has been classified by Belloni \etal \cite{bmk97} into twelve different X-ray states. 

The analysis presented here is based on the data reduction described
in full detail in Rau \& Greiner \cite{rg02}. We analyzed 139 observations from 89 days from the RXTE public archive of \grs\ from November 1996 to September 2000, when the source was in the $\chi$-state \cite{bmk97}. The absence of large amplitude variations and structured variability in these states suggests quasi-stable geometry and parameters during each observation and allows to fit the spectrum of an entire observation at once. $\chi$-states are connected with radio emission of varying strength and the most common states observed. 

From each observation we used data from PCU0 and HEXTE cluster 0 and fitted these with a model consisting of cold absorption (WABS), a disk blackbody (DISKBB) and a power-law spectrum reflected from an ionized relativistic accretion disk (REFSCH \cite{frs89},\cite{mz95}) in XSPEC \cite{a96}. Energies between 4 and 8.5\,keV were ignored during the fitting process because of the known response problems of the PCA at these energies. The hydrogen column density was fixed at $N_H$=5$\cdot$10$^{22}$\,cm$^{-2}$. 

Our model implies obvious simplifications, e.g. compared to the theoretical Comptonization models the power law overestimates the flux at low energies. Despite the simplicity of the three component model, we achieved surprisingly good fit results for our sample of $\chi$-state observations (reduced $\chi^2$$<$2 for all and $\chi^2$$<$1.1 for more than 60\% of the observations).

\vspace{-0.1cm}
\section{Comptonization of the X-ray radiation}
\vspace{-0.1cm}

Two main models for the origin of the hard power-law tail in X-ray binaries exist. i) the bulk motion Comptonization \cite{bp81},\cite{ct95}, where the soft disk photons are inverse Compton scattered on a cloud of free falling, hot electrons inside the last inner stable orbit around the black hole and ii) the disk corona geometry (e.g. \cite{hm93}), where the soft seed photons are scattered on a distribution of hot electrons located above the accretion disk.

The bulk motion model predicts a time lag between hard and soft X-ray photons of $\sim$0.2\,ms, much smaller than the observed time lags of 0.1\,s \cite{mrm01},\cite{rg02}. This rules out the bulk motion Comptonization as the origin of the hard X-ray photons in the $\chi$-states of \grs.  Therefore, the disk corona is the most probable geometry.  It is still unresolved if the hard power law tail continuous up to the MeV-range or if the spectrum cuts off. The electron distribution can either be thermal, non-thermal or hybrid. The absence of a clear cut off in the {\it RXTE} spectra however, requires high coronal temperatures ($\sim200$\,keV) and an optical depth, $\tau_T\ll 1$.

\begin{figure}[htb]
\vspace{-0.2cm}
\centering
\psfig{file=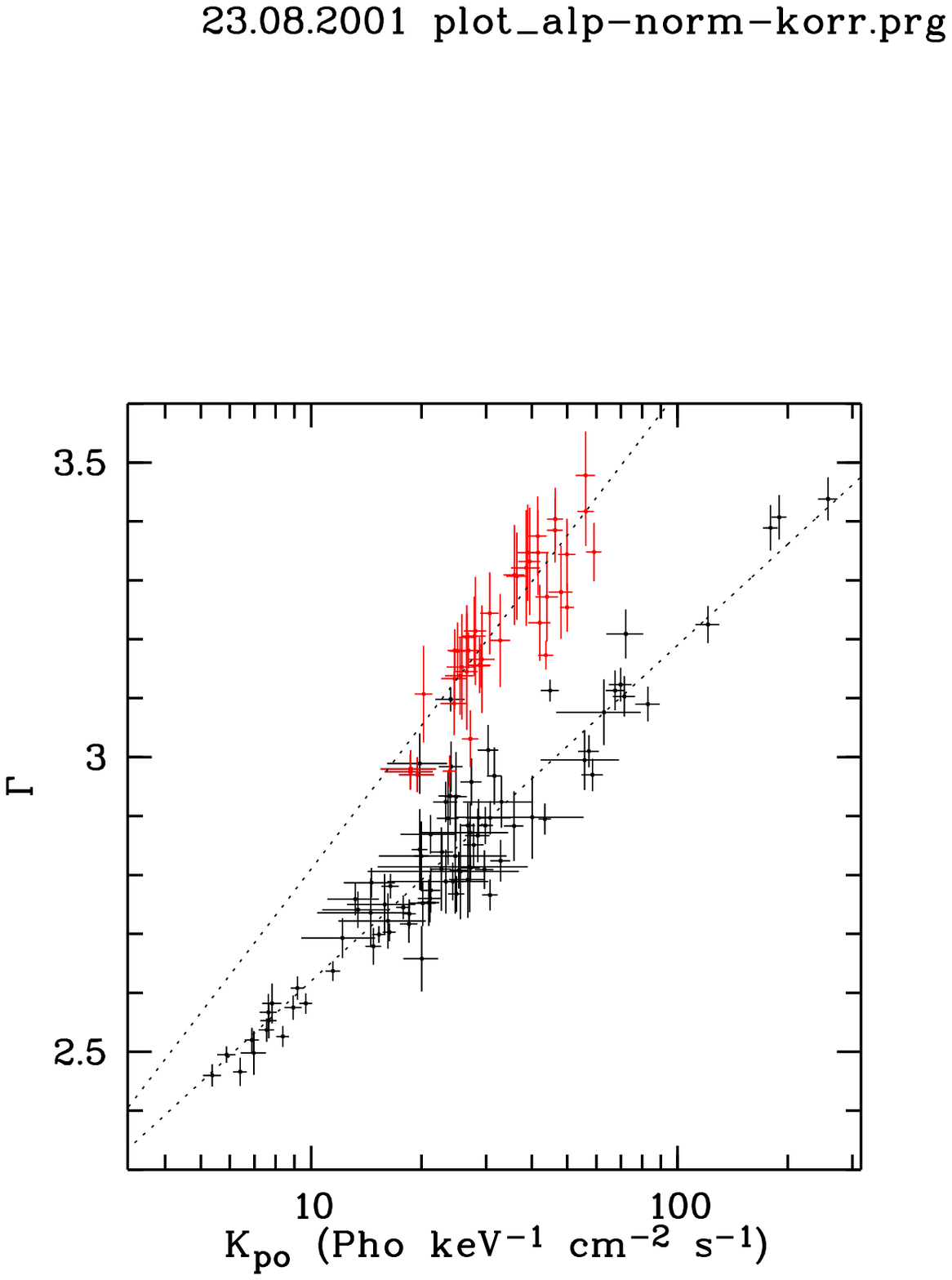,width=9.5cm,%
       bbllx=1.5cm,bblly=9.8cm,bburx=19cm,bbury=21.5cm,clip=}
\vspace{-0.2cm}
\caption{Power law slope, $\Gamma$, as a function of power law
          normalization, $K_{po}$. Each {\it RXTE} observation is
          represented by one data point. (dotted lines: correlation function)}
\label{fig:alpK}
\vspace{-0.5cm}
\end{figure}

The analysis shows a bimodal distribution of an increasing normalization, $K_{po}$, with increasing slope, $\Gamma$, of the power law (Fig.~\ref{fig:alpK}), suggesting a pivoting of the hard X-ray tail.  Pivoting was recently found also in Cyg X-1 \cite{zpp02} and can be explained by a variation in the flux of the soft seed photons or in the size of the corona. Assuming constant $\tau_T$, if the amount of intercepted soft seed photons is high, the electrons in the corona are cooled efficiently and the hard X-ray spectrum is steep. On the other hand, if the soft seed flux decreases, the temperature of the corona increases and the spectrum hardens. 

The two branches in Fig.~\ref{fig:alpK} represent observations during which long radio flares occurred (upper) and observations of times with short radio flares or without radio emission (lower) \cite{rg02}. Therefore, the different Comptonization behavior results probably from a difference in the optical depth or in the electron composition (thermal, hybrid).

\vspace{-0.1cm}
\section{Correlation of X-ray and radio emission}
\vspace{-0.1cm}

\begin{figure}[htb]
\centering
\psfig{file=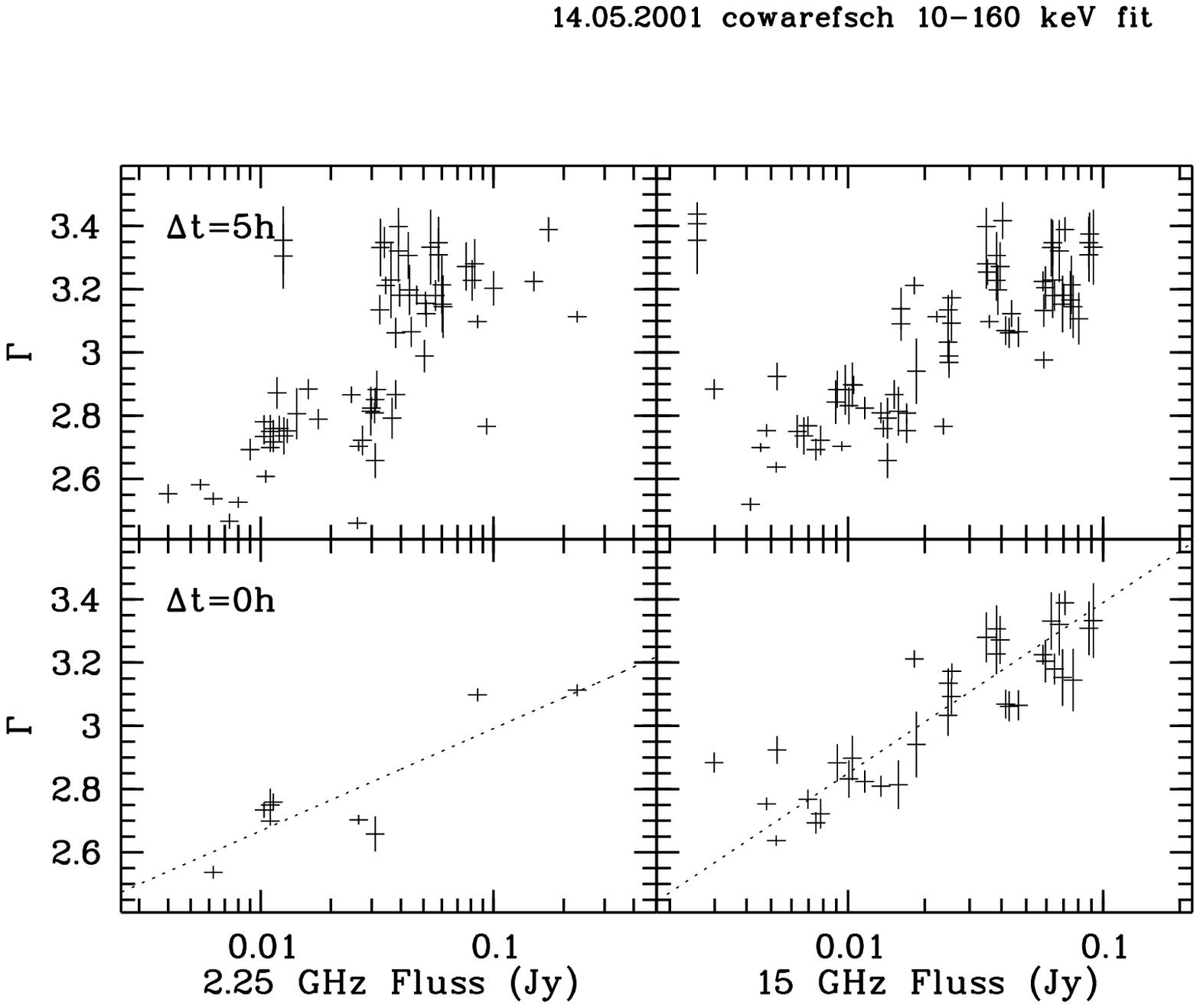,width=9.0cm,%
       bbllx=1.3cm,bblly=14.0cm,bburx=19cm,bbury=20.35cm,clip=}
\vspace{-0.3cm}
\caption{$\Gamma$($F_R$)-correlation for {\it GBI} (left) and {\it RT}
          (right) (dotted lines: correlation function)}
\label{fig:alphaRadio}
\vspace{-0.5cm}
\end{figure}

We analyzed 39 {\it Ryle Telescope, RT,} (15\,GHz) and 9 {\it Green Bank Interferometer, GBI}, (2.25 \& 8.3\,GHz) observations taken simultaneously with the RXTE observations and searched for correlations in all model parameters and fluxes. No correlation is seen between the soft X-ray and radio components. Instead, the slope of the hard power law tail correlates with the radio flux, $F_R$, (Fig.~\ref{fig:alphaRadio}).  

It is generally assumed that both the radio jets \cite{fg01}, \cite{f01} and the hard spectral component originate near the black hole. Similar to the interpretation of the pivoting behavior of the X-ray spectra, an increasing outflow of matter implies an increasing size of the scattering medium and therefore an increasing amount of intercepted soft seed photons. This leads to a lower plasma temperature due to cooling and to a softer X-ray spectrum.

This suggest the following scenario: when the radio emission is weak or absent, the Comptonization takes place in the corona above the accretion disk. With increasing outflow (higher radio emission) the matter in the jet intermingles with the corona. At high outflow rate the corona is blown away and the soft seed photons are scattered on the base of the jet \cite{f01}, \cite{rg02}. 

\vspace{-0.1cm}
\section{Long term periodicity}
\vspace{-0.1cm}
The analysis of the X-ray model parameters for all $\chi$-state
observations of \grs\ reveals a remarkably periodic behavior of
the hard X-ray component, for example the power law slope (see Fig.~4 in \cite{rg02}). An analysis of variance for  several model parameters and X-ray and radio fluxes was performed using the Fisher-Snedecor distribution function \cite{s89}. It shows a maximum at a period of 590$\pm$40 (FWHM)\,days in the hard X-ray components and in the radio fluxes of {\it RT} and {\it GBI} \cite{rg02b}. This confirms the 19\,month periodicity  which was already briefly mentioned by Kuulkers \etal \cite{kpk77}. 

\begin{figure}[htb]
\begin{center}
\psfig{file=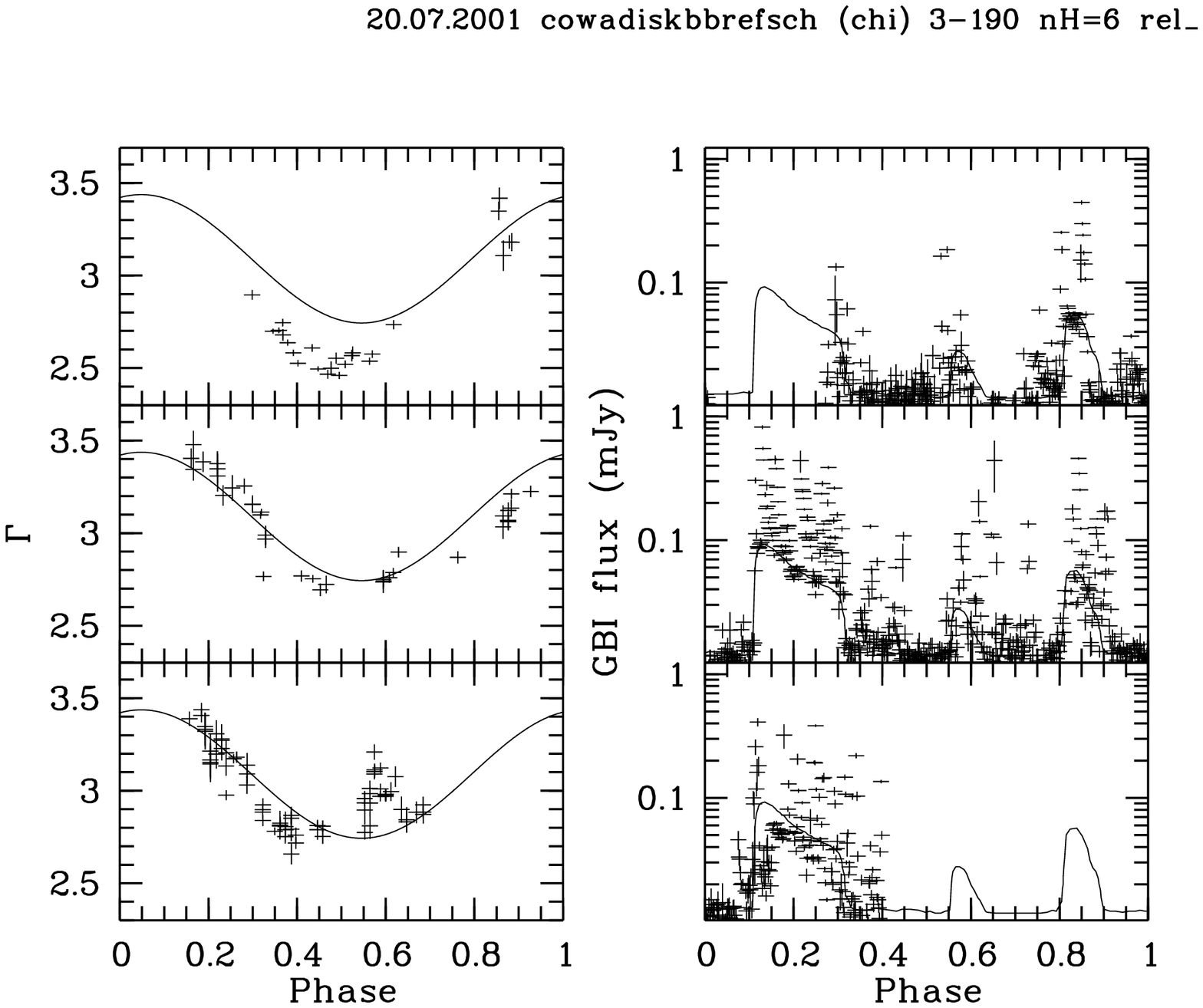,width=8.0cm,%
          bbllx=1.1cm,bblly=10.7cm,bburx=19.5cm,bbury=24.5cm,clip=}
\vspace{-0.25cm}
 \caption{Power law slope of 139 $\chi$-state observations (left) and 2.25\,GHz radio flux (right) from JD 2450350--2450940, JD  2450940--2451530 and JD 2451530--2452120 (top to bottom) in periods of 590\,days. The solid lines visualize the best fitting sinusoidal function (left) and illustrate the repeating outbursts in the radio flux (right).}
\end{center}
\label{fig:phasePho}
\vspace{-0.7cm}
\end{figure}

Fig.~\ref{fig:phasePho} shows the power law slope (left) and the {\it GBI} radio flux at 2.25\,GHz in periods of 590\,days. Where $\Gamma$ has a quasi-sinusoidal behavior the periodicity in the radion flux is visible in the radio outbursts. Both the broad outburst with exponential cutoff at phase $\sim$0.2 and the shorter and steeper outbursts repeat periodically. 
  
Long-term periodicity in X-ray binaries is known since more then two decades \cite{k73} and radiation driven warping of the accretion disk is the generally assumed mechanism for the periodicity in low-mass X-ray binaries \cite{p77}, \cite{od01}. The outer, tilted part of the
accretion disk is irradiated by the central source. If the radiation is
absorbed and re-emitted parallel to the local disk gradient, a torque due to gas pressure affects the disk. A small warp grows exponentially if the luminosity is sufficient to depress the dissipative processes forcing the disk back into the orbital plane \cite{p77}. 

The periodicity accrues because the Roche lobe overflow of the donor star hits the not co-rotating accretion disk at different disk radii during an orbital cycle. This leads to regions in the disk with differing mass density. The radial flow in the inner part of the accretion disk is therefore not
homogeneous leading to a periodic variation of the amount of soft seed
photons being comptonized in the corona. Therefore, the electron
temperature in the corona varies on the same time scale due to Compton
cooling and the power law slope varies. The amount of matter to
be ejected in a jet changes with similar periodicity which explains
the overall periodicity in the radio flux.  

The soft X-ray component (ASM count rate \& hardness ratios, soft model components) do not show any periodicity in the $\chi$-state observations. The periodicity can simply be masked by the high short term variability in timing and spectral behavior of \grs. 

\vspace{-0.1cm}
\section*{Acknowledgments}
\vspace{-0.1cm}
The authors thank M. L. McCollough (BATSE) and
G. G. Pooley (MRAO) for providing the BATSE and {\it RT} data. {\it The Ryle Telescope} is supported by PPARC. The {\it Green Bank Interferometer} is a facility of the National Science Foundation operated by the NRAO in support of NASA High Energy Astrophysics programs.

\end{document}